\documentclass[preprint,proceedings]{rmaa}

\suppressfulladdresses 
\usepackage{paralist}
\usepackage{psfrag,color}

\SetYear{2006}
\SetConfTitle{XI Latin-American Regional IAU Meeting}

\title{Sub-Structures in the Halo of the Milky Way} 

\author{A. Katherina Vivas\altaffilmark{1}, Robert Zinn\altaffilmark{2},
Yolimar Subero\altaffilmark{1} and Jes\'us Hern\'andez\altaffilmark{1}}
\altaffiltext{1}{Centro de Investigaciones de Astronom\'{\i}a (CIDA),
Apartado Postal 264, M\'erida 5101-A, Venezuela (akvivas@cida.ve, ysubero@cida.ve,
jesush@cida.ve).}
\altaffiltext{2}{Yale University, Department of Astronomy, PO Box 208101,
New Haven, CT 06511, USA (zinn@astro.yale.edu).}

\shortauthor{Vivas et al.}
\shorttitle{Sub-Structures in the Halo}

\listofauthors{A. K. Vivas, R. Zinn, Y. Subero \& J. Hern\'andez}
\indexauthor{Vivas, A. K.}
\indexauthor{Zinn, R.}
\indexauthor{Subero, Y.}
\indexauthor{Hern\'andez, J.}

\abstract{
The latest results of the QUEST survey for RR Lyrae stars are described.  This survey, 
which is designed to find and characterize sub-structures in the halo of our Galaxy, has
covered about 700 square degrees of the sky and has detected 693 RR Lyrae stars, most 
of which are new discoveries.  The spatial distribution of the RR Lyrae stars reveals several
interesting groups in the halo.  Some of them appear to be related to previous detections 
of the destruction of dwarf spheroidal galaxies (the Sagittarius tidal streams, the 
Virgo stellar stream, the Monoceros ring) or globular clusters (Palomar 5), while others 
still have an unknown origin.}

\resumen{Se describen los \'ultimos resultados de la b\'usqueda QUEST de 
estrellas RR Lyrae. Esta b\'usqueda, la cual fue dise\~nada para encontrar y 
caracterizar sub-estructuras en el halo de nuestra Galaxia, ha cubierto aproximadamente 700
grados cuadrados del cielo y detectado 693 estrellas RR Lyrae, la mayor\'{\i}a
de ellas descubiertas por primera vez en este estudio. 
La distribuci\'on espacial de las estrellas RR Lyrae
ha revelado varios grupos interesantes en el halo. Algunos de ellos parecen
estar asociados con la destrucci\'on de galaxias esferoidales enanas 
(las colas de marea de Sagitario, la corriente estelar de Virgo, el anillo de Monoceros)
o c\'umulos globulares (Palomar 5), mientras que otros grupos tienen todav\'{\i}a un
origen desconocido.}

\addkeyword{Galaxy: structure}
\addkeyword{Galaxy: halo}
\addkeyword{Stars: variables: others}

\begin{document}
\maketitle

\section{Introduction}

Over the past few years, several detections of debris from disrupted galaxies 
have been made by surveys of the halo of the Milky Way
\citep[][among others]{new02,maj03,yan03,viv06}. 
Some of them are described elsewhere in 
this Volume (see contributions by S. Duffau and H. Rocha-Pinto). The debris is 
usually observed as an excess 
in the number density of different stellar tracers (e.g. M giants, F-type main-sequence stars,
horizontal branch stars), and/or as groups of stars with similar kinematic properties.
These results indicate that the density distribution of the stars in the Galactic halo 
is not smoothly varying, as was once believed.  Ancient mergers with small satellite galaxies seem to have left a signature of intricate streams, as predicted by some theoretical models
\citep{bul01,bul05}.
Some small sub-structures may have been produced instead by the destruction of globular clusters \citep{ode03,gri06}. 
To find those tidal streams and characterize their stellar 
populations are important for the understanding of the formation of galaxies
like the Milky Way, and for constraining the theoretical models of hierarchical
galaxy formation.

We present here the QUEST survey for RR Lyrae stars which is aimed to study the spatial
distribution of this type of horizontal branch star in the halo. RR Lyrae stars
are expected to be excellent tracers of tidal streams not only because they are good
standard candles (hence, groups in the 3D halo are pinned down reliably) but also
because they have been observed in all of the dwarf spheroidal (dSph) 
galaxies around the Milky
Way \citep[see][]{viv06}. It is believed most mergers have involved that type of small galaxies 
(but see contribution by M. Catelan in this volume), and thus, presumably all streams
contain a population of RR Lyrae stars.

\section{The QUEST Survey}

The QUEST survey for RR Lyrae stars is being carried out with the 1m Jurgen Stock 
Telescope at the Venezuelan National Observatory of Llano del Hato.
It uses the QUEST camera \citep{bal02}, 
an array of 16 CCDs, in driftscan mode. We repeately scan
strips of the sky of constant declination in several photometric bands. 
We use the V-band observations
as the main bandpass for constructing time-series and recognizing RR Lyrae stars.
The complete description of the survey and algorithms used for detecting RR Lyrae stars
can be found in \citet{viv04}.

\subsection{Observed Region}

\begin{figure}
  \includegraphics[bb=17 145 591 494,width=\columnwidth]{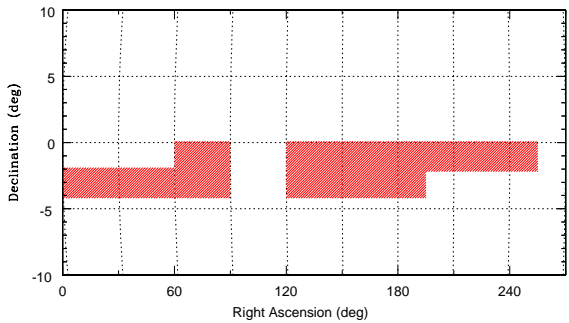}
  \caption{Region covered by the QUEST survey for RR Lyrae stars
in equatorial  coordinates.}
\end{figure}

To date, we have observed and analyzed $\sim 700$ square degrees
of the sky, in two strips of constant declination near the
celestial equator. The region observed is shown in Figures 1 and 2 in
equatorial and galactic coordinates respectively. 
Since our main goal is to study the Galactic halo, we skipped a region near 
the galactic plane (around RA$\sim 7$h
at this declination).  Nonetheless, the survey covers a wide range in galactic latitude.
The data corresponding to the strip centered on 
$\delta=-1^\circ$ has already been published \citep{viv04}.

\begin{figure}[t]
  \includegraphics[width=\columnwidth]{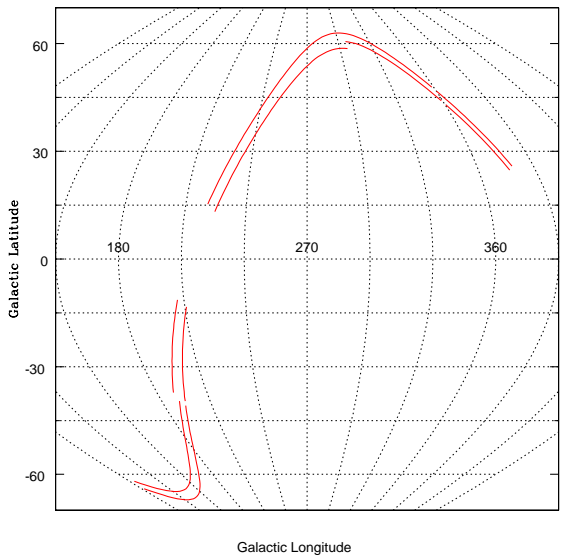}
  \caption{Region covered by the QUEST survey in galactic coordinates.}
\end{figure}

\subsection{Known Problems}

The custom-made automated software for the detection of objects in the QUEST data 
has no capabilities for deblending close objects. 
We have discovered that blended objects may appear variable because of seeing 
variations and that several of the stars in the first QUEST catalog are not real 
RR Lyrae variables but blended objects.  Most of these blends occurred in the regions 
of the lowest galactic latitude where crowding is more severe, which is also where the 
fewest number of observations had been made.  With relatively few epochs available for 
each star, there was a greater chance that a fake variable might mimic the light curve 
of a RR Lyrae star and therefore be included in the QUEST catalog.  \citet{viv06} have
eliminated all known blends from the catalog.  
At high galactic latitudes, only ~1\% of the variables were cases of blended images.

Another source of contamination in the first catalog was found after spectroscopic 
follow-up.  We have found a few cases where the spectra of type c stars indicate 
much cooler than expected effective temperatures. Type c stars are
always located in the blue side of the instability strip.  These unusually cool type c variables are probably instead variable blue stragglers or W UMa eclipsing variables.  
Again, this type of contaminating object
appears mostly at low galactic latitudes \citep[see][]{viv04}, and only for type
c stars which are less common than the types ab.

\section{Spatial Distribution of RR Lyraes in the Halo}

\begin{figure*}[!t]
  \includegraphics[width=\columnwidth]{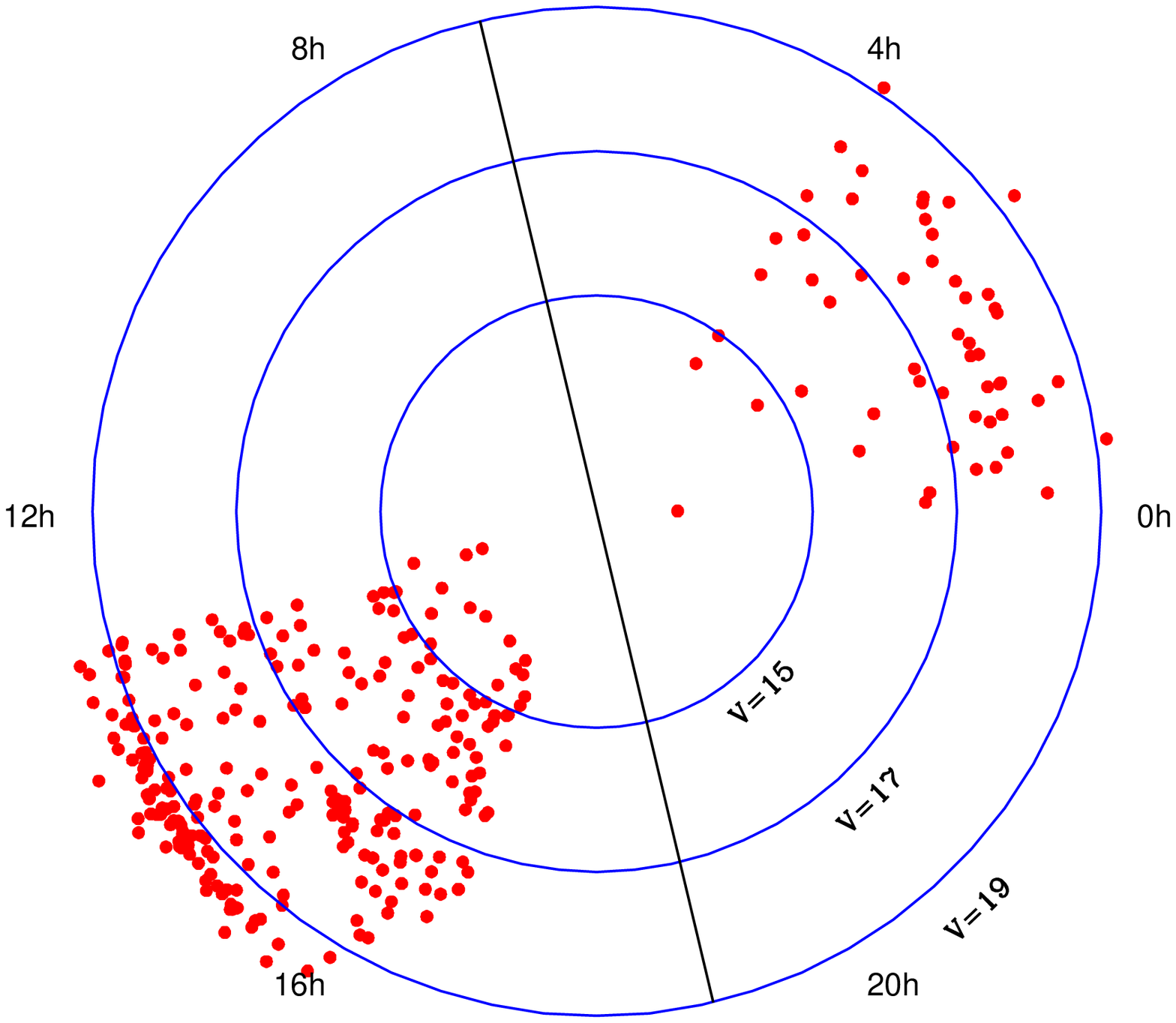}
  \hspace*{\columnsep}
  \includegraphics[width=0.95\columnwidth]{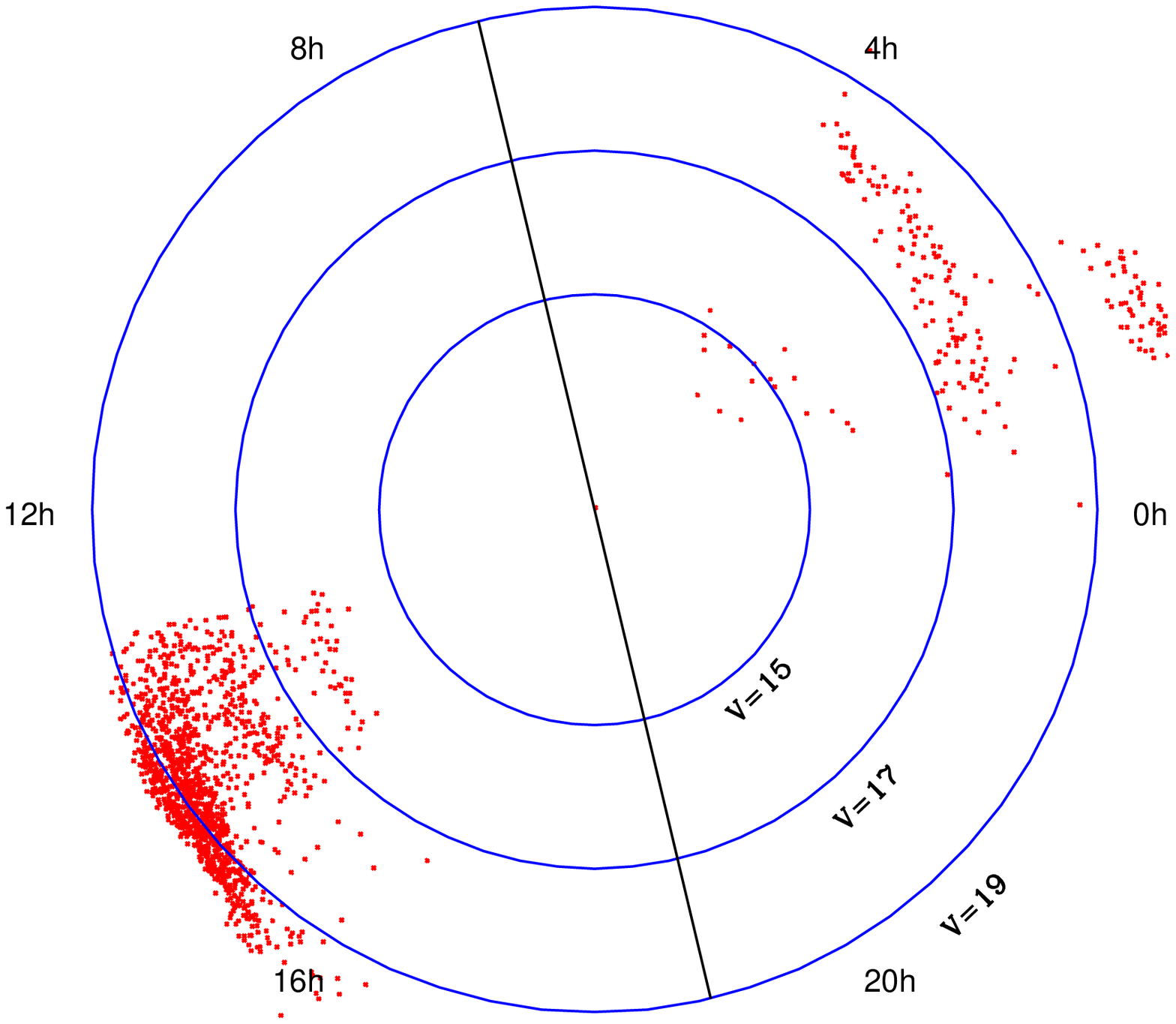}
  \caption{Left: Distribution of the RR Lyrae stars in a slice of 2.2 deg wide.
The region is, by coincidence, dominated by debris from the Sgr dSph galaxy.
In the right plot we show the distribution of such debris as predicted by the
numerical simulations of Law te al. (2005). The simulation shown assumes 
a spherical potential for the Milky Way's dark matter halo. In both plots, the circles
correspond to a distance from the Sun of 8, 19 and 49 kpc.}
\end{figure*}

Figures 3 shows the spatial distribution of the QUEST RR Lyrae stars in the
regions where a single declination strip was surveyed. Thus, the observed region
is a 2.2 deg wide strip of the sky. The radial axis in these plots correspond
to extinction corrected V magnitudes, which is equivalent to distance from the
Sun for RR Lyrae stars. 
The bright and faint limits of the survey enable the detection of RR Lyrae stars 
from $\sim 4$ and $\sim 60$ kpc from the Sun. The spatial
distribution of the RR Lyrae stars (see also Figure 4) shows that the halo does not
have a smooth distribution. Several sub-structures are present.

In particular, the Sagittarius (Sgr) tidal streams are seen in Figure 3. The plot in the
right shows Sgr stream particles from the numerical simulations by \citet{law05}
in exactly the same region of the sky as our observations. A very
strong overdensity is located at $\sim 50$ kpc from the Sun, at
$13 < \alpha < 16$ h. This is part of the Sgr leading tail, and it is clearly 
seen in our data (left plot). According to the model, there is also Sgr debris (part of 
the trailing tail) 
at $\sim 28$ kpc, between $0 < \alpha < 4$ h. The number density of this part of the 
trailing tail of Sgr is about 1/10 of the density in the leading tail. This is
the reason why it is not as obvious in the QUEST data. However, a 
more detailed look in our data indicates
that there is indeed a peak in the number density of RR Lyrae stars at $V=17.8$, 
which agrees with
both the previous observations of this part of the tails with
other tracers \citep{new02,maj03}, and the Sgr models by \citet{law05}, 
at least in the versions assuming 
either an spherical (Figure 3, right) or prolate shape of the dark matter halo of the
Milky Way. The model assuming an oblate shape predicts the trailing tail at a closer
distance. Measurements by \citet{viv05} of the spatial and velocity distributions of the 
QUEST RR Lyrae stars in the leading tail of Sgr have also indicated that spherical and 
prolate models provide better fits to the data.

The QUEST data shown in Figure 3 includes also the globular cluster Palomar 5 (at
$\alpha \sim 15\fh 3$, $V=17.3$. Not only we recovered the 5 known RR Lyrae variables
in this cluster, but we also discovered 2 new members. In addition, we detected a strong
excess of RR Lyrae stars in the region surrounding the cluster. Most likely, these stars
are related with the cluster which is known to be in the process of
tidal disruption \citep{ode03}. In \citet{viv06}, we modeled the galactic 
background of RR Lyrae stars and used it to
detect overdensities in the first catalog. 
The significance of these results was investigated by using Monte Carlo simulations 
to find the likelihood that random variations alone can produce overdensities that 
are similar in size to the ones observed.
Both the Sgr leading tail and the stars around Pal 5 appear to be significant overdensities.

\begin{figure}[!t]
  \includegraphics[width=\columnwidth]{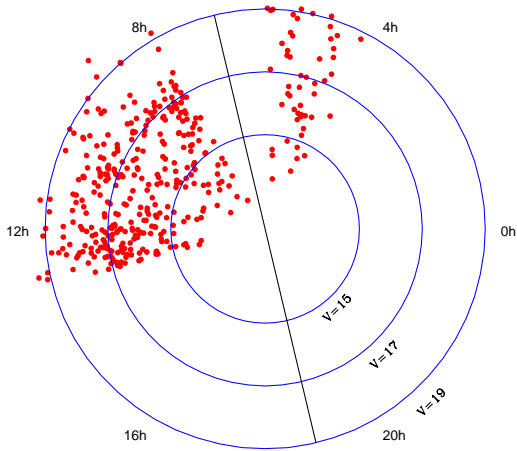}
  \caption{Distribution of RR Lyrae stars in a slice of 4-deg wide. The
big overdensity at $\sim12$ hours of RA is due to the Virgo Stellar Stream
and probably another sub-structures.}
\end{figure}

In Figure 4, we show a slice of the sky that is $4^\circ$ wide in declination 
(includes two QUEST strips).  Because more stars are included than in the $2\fdg 2$ 
slice of the first QUEST catalog, the presence of sub-structure is more evident.

The strongest excess of RR Lyrae stars is seen at $12<\alpha <13$h and $V\sim 17$, which
is equivalent to a distance of $\sim 20$ kpc from the Sun. This feature was observed
before with just the first QUEST catalog 
\citep{viv01,viv03}, and in data from SDSS \citep{new02}. 
Spectroscopic observations of a sample of RR Lyrae stars in this region have suggested 
that part of this overdensity is the remnant of a disrupted dSph galaxy
(see contribution by S. Duffau in this Volume, and Duffau et al.
2006). The newest QUEST data (Figure 4) indicates that the feature extends to the south
and continues being a equally strong overdensity in the southermost part. Investigation
of the radial velocities and metallicities of the rest of the stars in the region continues.

Figure 4 shows other sub-structures. A couple of them may have a common origin, the Monoceros ring \citep{yan03,iba03}. Our surveyed region crosses the Monoceros ring in two parts: at $\alpha \sim 4\fh 8$, $V\sim 15.5$, and at $\alpha \sim 8\fh 5$, $V\sim 15.7$.
Both features were detected by our overdensity finding algorithm \citep{viv06}, using
only stars in the first catalog.

With the addition of the second QUEST strip, a new feature is now clearly visible at
$\alpha \sim 10\fh 5$, $V\sim 17.3$. It contains about 15 RR Lyrae stars closely located
in the sky, at a mean distance of $\sim 23$ kpc from the Sun. Another possible large group 
is also seen at a similar magnitude, between $8<\alpha < 9$h. Spectroscopy of stars in
these groups is needed to confirm that they are indeed a coherent group in velocity space, 
and to investigate their origins.

\section{Concluding Remarks}

The results obtained so far by the QUEST survey indicate that RR Lyrae stars
are powerful tracers of sub-structures in the halo. This is due in part to the
fact that they are the best standard candles among the commonly used 
tracers of the halo population. The determination of radial velocity and metallicity
distributions in the halo sub-structures is needed to understand the origin of the
overdensities and to build a better picture of the accretion history of the Milky
Way.

The QUEST survey will continue until covering $\sim 1200$ sq degrees of the sky. It
will include a study at low galactic latitudes toward the Galactic anticenter in order to investigate the number density and extension of the thick disk population of RR Lyrae stars.
In addition, a database containing the QUEST timeseries for several 
tens of thousands objects will be made available to the astronomical community.

\acknowledgements

The Llano del Hato Observatory is operated by CIDA for 
Ministerio de Ciencia y Tecnolog{\'\i}a of Venezuela.
This research was partially
supported by the National Science Foundation under grant 05-07364.
AKV thanks FONACIT (Venezuela) and the IAU for providing funding to attend 
this conference.

\end{document}